\documentclass[a4paper]{jpconf}
\usepackage{graphicx}
\usepackage{eurosym}
\usepackage{amssymb}
\usepackage{amsmath}
\usepackage[symbol]{footmisc}
\usepackage{upgreek}
\renewcommand{\thefootnote}{\fnsymbol{footnote}}

\begin{document}
\title{Precision laser diagnostics for LUXE}

\author{Rajendra Prasad for the LUXE collaboration}

\address{FTX, Deutsches Elektronen-Synchrotron DESY, Notkestr. 85, Hamburg, 22607, Germany}

\ead{Rajendra.Prasad@desy.de}

\begin{abstract}
Strong field QED is an active research frontier. The investigation of fundamental phenomena such as pair creation, photon-photon and photon-electron interactions in the nonlinear QED regime are a formidable challenge – both experimentally and theoretically. Several experiments around the world are being planned or in preparation to probe this strong field regime.

LUXE (Laser Und XFEL Experiment) is an experimental platform which envisages the collision of the high quality 16.5 GeV electron beam from the European XFEL accelerator with a 100 TW class high power laser. One of the unique features of LUXE is to measure the key observables such as pair rates ($e^+ e^-$) with unprecedented accuracy in the characterization of both beams together with ample statistics. The state-of-art detector technologies for high energy particle/ photon detection enable percent level precision. The state-of-art high power lasers offer high quality laser beams, however, the residual shot-to-shot fluctuations coupled with the large nonlinearity of the processes under investigation form a particular challenge. An uncertainty of $5 \%$ on the absolute laser intensity already leads to a very large ( about $40 \%$) uncertainty in the pair rate. Hence it becomes essential to control the laser parameters precisely.

To mitigate this issue a full suite of laser diagnostics is being currently developed at the JETI40 laser in Jena with the aim of tagging the shot intensity to $<1 \%$. In this presentation, details of the laser and the diagnostics suit for the single shot tagging of all the laser parameters will be presented. Moreover, results from an ongoing campaign to properly relay image the beam without significant distortion of the laser beam parameters for post-diagnosis will be discussed. 
\end{abstract}

\section{Introduction}
The quantum electrodynamics (QED) is  the most precise tested theory in modern physics. It has stood the test of experimental precision by one part in billion for the case of electron anomalous moment, for example \cite{Hanneke2011}. The backbone of the QED prediction power is calculations based on perturbative method. However, a limit appears in the vicinity of  a super strong electromagnetic field. Above the so-called critical field number of novel phenomena can occur such as vacuum polarisation, pair production from vacuum, to name a few, and a non-perturbative approach has to be employed to correctly interpret these processes. The critical field is described by following expression \cite{PiazzaRMP}.
\begin{equation}{\label{eq:critical_field}}
    E_{cr}= \frac{m^2_e c^3}{e \hbar}\approx 1.3\times 10^{18}  \; \mathrm{V/m} ,
\end{equation}
where $e$ and $m_e$ are elementary charge and rest mass of the electron, $c$ is the speed of light, and $\hbar$ is the reduced Planck's constant.

\begin{figure}[ht]
\includegraphics[width=19pc]{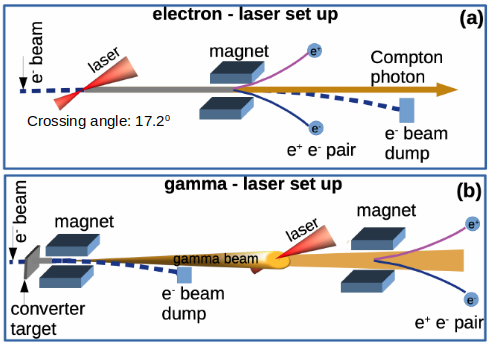}
\includegraphics[width=17.5pc]{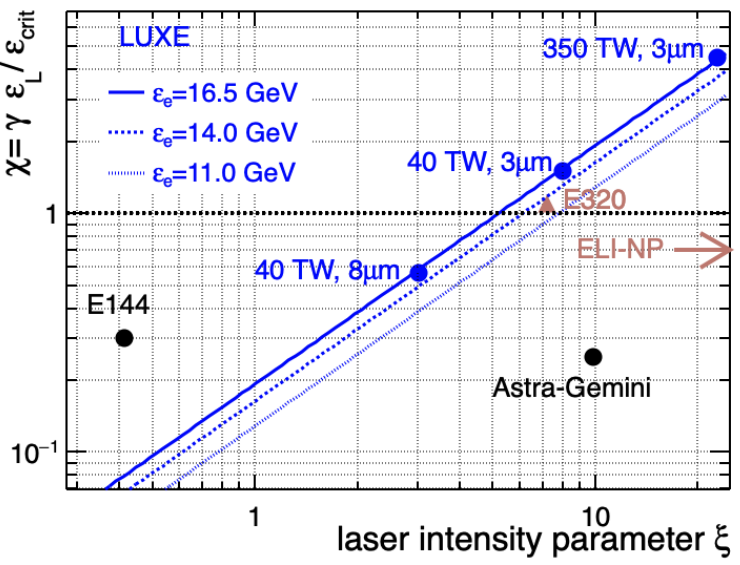}
\caption{Left panel: two distinct experimental configurations to probe the nonlinear QED regime (a) in  electron laser collision ``e-laser" mode and (b) in gamma photon and laser collision ``g-laser" mode. Right panel: Parameter space covered by LUXE experiment. The normalised laser amplitude $\xi$ is reached up to $>23$ leading to accessible $\chi$ values $>4$. Three blue lines correspond to different electron beam energies: 11.0 GeV (fine dotted line), 14.0 GeV (dotted line ) and 16.5 GeV (solid line). Two blue solid points labelled as ``40 TW, 8$\upmu$m" and ``40 TW, 3$\upmu$m" show the laser power and focal spot size for Phase 0, the third one labelled as ``350 TW, 3$\upmu$m" represents the  Phase 1, which will be achieved after the upgrade of the laser system.}  
\label{elaser-glaser}
\end{figure}

At present such fields are not directly available in the laboratories. The state-of-art high power lasers now a days can reach focused intensity in excess of $10^{21} \mathrm{W/cm^2}$, which can enable fields of the order of $10^{14} \mathrm{V/m}$. However, this  still falls short by more than three orders of magnitude. In order to reach the critical field, LUXE \cite{Abramowicz2021} envisages to collide the high quality electron beam of EuXFEL with high power laser to boost (relativistic effect) the electric field in the rest frame of electron. In Fig. \ref{elaser-glaser} we show two experimental configurations that are planned to explore all the physics processes.

In Fig. \ref{elaser-glaser}(a), the so-called e-laser mode is illustrated. The high energy (up to 16.5 GeV) electron beam collide with a high intensity laser. The main physics process which will be studied here are nonlinear Compton scattering and pair production via nonlinear Breit-Wheeler process \cite{Breit1934,Reiss1971, Hartin2019,Blackburn2018}, which are given by equations \ref{eq:nl_CS} and \ref{eq:BW_elaser}. 
\begin{equation}\label{eq:nl_CS}
    e^- + n \gamma_L \longrightarrow e^- \gamma_c ,
\end{equation} 

\begin{equation}\label{eq:BW_elaser}
    \gamma_c + n \gamma_L \longrightarrow e^+ e^- ,
\end{equation} 
where $\gamma_L$ and $\gamma_c$ represent laser and inverse Compton photons and $n$ is the number of photons involved.
The second configuration is shown in Fig. \ref{elaser-glaser}(b). Here the high energy photons will be produced using two different mechanisms. In the first case, a converter target will be used to produce high energy photons via bremsstrahlung. The second case will employ an inverse Compton scattering method to produce high energy photons with a narrower energy spectrum. The produced photons in turn then will collide with the high intensity laser pulse. The uniqueness of this geometry is that it will enable a pure photon-photon collision in the nonlinear regime, which has not been experimentally achieved thus far. The main physics process explored here is pair production via nonlinear Breit-Wheeler process:
\begin{equation}\label{eq:BW_glaser}
    \gamma_B + n \gamma_L \longrightarrow e^+ e^- ,
\end{equation} 
  
where $\gamma_B$ refers the bremsstrahlung photon. 

The key parameters in all these studies  are the so-called quantum nonlinearity parameter $\chi$ and the normalised amplitude of the laser $\xi$ (indicator of laser intensity ). They are defined by the following expressions:
\begin{equation}
    \xi= \frac{e E}{m_e \omega c}\;\; \mathrm{and} \; \;\; \chi = \frac{E^*}{E_{cr}} ,
\end{equation} 
where $E$ is the amplitude of the laser electric field, $\omega$ is the angular frequency and $E^*$ is the boosted electric field seen by the electron in its rest frame.

The parameter space covered at LUXE is depicted in the right hand panel of Fig. \ref{elaser-glaser}. The experiment will be performed in a phased manner. Initially in phase 0, a 40-50 TW laser will be installed. This will enable exploration of the nonlinear regime with  $\chi \simeq 1$. In phase 1, an upgrade of laser to 350 TW will expand the explored parameter space to $\chi > 4$. Thus a vast range of scans ranging from $\xi=0.5-23$ will be covered enabling a clear identification of transition from perturbative to non-perturabtive regime, for instance. In addition, the first experiment studying the perturbative transition in the collision of electron beam with laser was performed at SLAC \cite{E144, E144-2} is also marked (black solid points, labelled as ``E144"). The parameter range covered in  this experiment was mostly in the perturbative regime. We also show the purely laser-based, all-optical schemes that have been employed at Astra-Gemini \cite{Cole2018, Poder2018} (black solid point, labelled as ``Astra-Gemini") and are planned at, e.g. the Extreme Light Infrastructure (ELI) \cite{Weber2017, Gales2018}.

\section{Requirement of precision laser diagnostics}
One of the unique features of LUXE experiment is that the measurements are aimed to be performed with unprecedented accuracy and ample statistics. This poses stringent requirements on both the machines involved (electron and laser). The EuXFEL has  very high quality electron beam parameters.  However, the residual shot-to-shot fluctuations of laser parameters coupled with the large nonlinearity of the processes under investigation form a particular challenge. First we discuss the important role played by laser intensity on the key physics observable. 

High energy Ti:Sapphire lasers using flash-lamp technology for the pump-lasers have typical energy fluctuations of around 2--3\% rms. These overall energy fluctuations, while significant, are not the dominant contribution to shot-to-shot intensity fluctuations. The major contributions to such fluctuations are small changes in phase both in real space and in frequency space due to temperature drift, air currents and mechanical vibrations. Small changes in spatial phase result in the spot radius fluctuating at the few \% level resulting in up to 10\% level intensity fluctuations. Similarly, fluctuations in the spectral phase can lead to 1\% rms fluctuations in the pulse duration. In total, the variation in between the highest and lowest intensity shots  on a stable laser can reach $\sim$ 15\%.
Simulations show that an uncertainty of about $5\%$ in intensity can lead up to $40\%$ uncertainty in the pair rate. Hence, it become detrimental to know the laser intensity with very good accuracy. 

\section{Principle of precision diagnostics}
\begin{figure}[htbp]
\includegraphics[width=37pc]{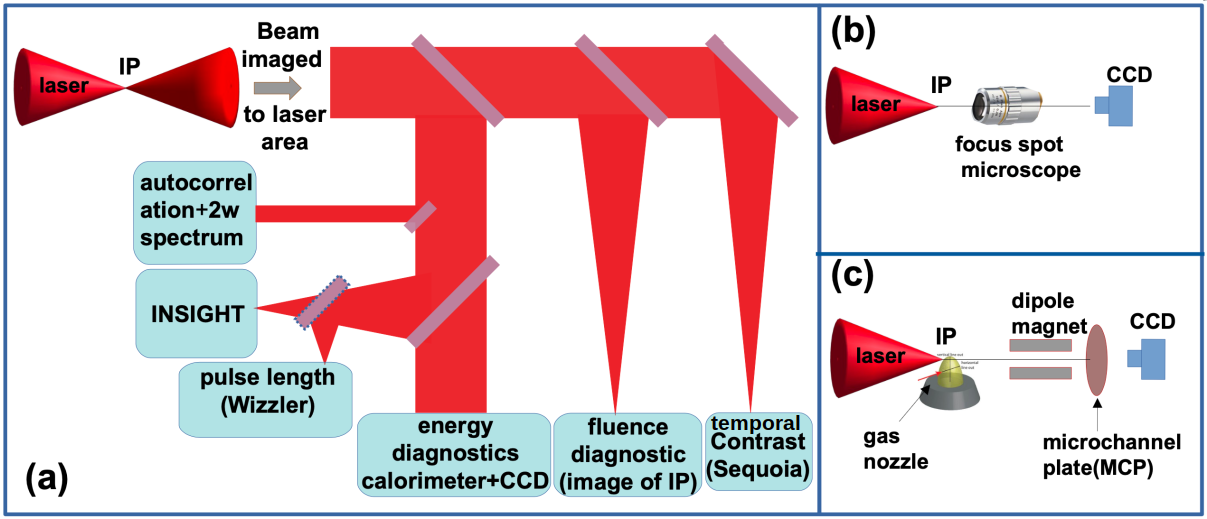}
\caption{Single shot tagging scheme to enable precision control of the laser parameters. (a) Tagging of primary laser parameters based on the relay imaging of the laser beam to the relevant diagnostics. (b) Focus diagnostic to optimise and measure the focus at the IP. (c) An indirect approach to measure the laser intensity.}
\label{tagging_scheme}
\end{figure}

The aim is to diagnose the laser intensity, on a shot-to-shot basis, with percent level accuracy. A dedicated state-of-art diagnostic system capable of measuring the fluctuations will be set up.  A schematic of the proposed diagnostic system is shown in Fig.~\ref{tagging_scheme}. The shots will then be tagged with their precise relative intensity allowing precision relative measurements using various diagnostics. A key factor allowing us to tag the laser parameters on shot-to-shot basis is the fact that during the interaction with electron bunch, the laser beam is not significantly perturbed or absorbed (in contrast to, for example, high-power laser interactions with solid or gas targets). Hence as shown in Fig.~\ref{tagging_scheme}, the laser beam will be relay imaged to laser area for tagging.
One of the key features will be imaging of the laser beam to ensure that the beam intensity profile at the compressor output, the focussing parabola at interaction point and at entry of the returning beam have the same phase and amplitude distribution, thus providing the highest possible fidelity of the diagnostic suite.  

The tagged parameters are primarily the laser pulse energy, pulse duration and focal spot size. In addition, we shall also use  INSIGHT diagnostics which offers full 3D characterisation of the focus and hence enables determination of the electric-field distribution in the focus at the interaction point (see Fig. \ref{tagging_scheme}(a)). We note that this device will be used offline and an average measurement will be taken. The focal spot size at the interaction point will be also checked using high magnification focus diagnostic (see arrangement in Fig. \ref{tagging_scheme}(b)). An indirect approach to measure the intensity at the IP is also illustrated in  Fig. \ref{tagging_scheme}(c). The method is based on ponderomotive scattering of the  free electrons \cite{FCC}. 

\section{Offline diagnostic testing}
The offline setup of the diagnostics and testing are already being performed at the JETI40 laser facility of Helmholtz Institute Jena \cite{Jeti40}. JETI40  is a Ti:Sa laser system operating at central wavelength of 800 nm. It delivers pulses at 10 Hz repetition rate. The energy in a single pulse can reach in excess of 1 J in a pulse duration of 30 fs (FWHM). The last amplification stage consists of a cryogenic cooled amplifier system. The beam is expanded to 5 cm and enters the grating based compressor. After the compressor, an adaptive optics apparatus enables wavefront correction of the beam and hence a diffraction limited focus spot can be obtained.

As discussed above, the proper imaging of the beam to the diagnostic suite is a very important aspect of the system.  The  relay imaging of the beam is based on telescope imaging.  Based on the calculations with reference parameters, two telescopes lines will be needed. We envisage using an off axis  parabola (OAP) in combination with an infinitely corrected microscope objective as first telescope. The second telescope will be based on 4f relay, with no change in the magnification, i.e, 1:1 imaging. The setup is in primary phase of testing. 

\section{Summary}
The LUXE experimental platform presents an exciting opportunity to probe the strong field QED in uncharted territory. The parameter space covered at LUXE enables studies of some novel phenomena proposed by theory such as the intensity-dependent position of the nonlinear Compton edges and the transition from the perturbative to nonperturbative regimes in electron-positron pair production. The test of theoretical predictions relies on accurate measurements of key physics observables. In terms of the particle and photon detection aspects, enabling this goal demands the development of a state-of-art detector suite. On the other hand from the laser side, even more stringent requirements on its parameters demands state-of-art laser diagnostics that can precisely control and monitor the relevant pulse parameters. The single shot tagging method described above aims to tag the shot to shot intensity with $<1 \%$ level of accuracy. 
\section{Acknowledgments}
Author acknowledges the "Bundesministerium f\"ur Bildung und Forschung" for the support via the 2018 Helmholtz Innovation Pool. This work has also benefited from computing services provided by the German National Analysis Facility (NAF).


\section*{References}
\begin{thebibliography}{99}
\bibitem{Hanneke2011} D. ~Hanneke, S. ~Fogwell Hoogerheide and G. ~Gabrielse 2011 Cavity control of a single-electron
quantum cyclotron: Measuring the electron magnetic moment {\it Phys. Rev. A} \textbf{83} 052122 
\bibitem{PiazzaRMP} A. ~Di Piazza, C. ~M\"uller, K. Z. ~Hatsagortsyan and C. H. ~Keitel 2012 Extremely high-intensity laser interactions with fundamental quantum systems {\it Rev. Mod. Phys.} \textbf{84} 1177
\bibitem{Abramowicz2021} H. ~Abramowicz {\it et al.} 2021 Conceptual Design Report for the LUXE Experiment {\it Eur. Phys. J. ST} \textbf{230} 2445
\bibitem{Breit1934} G. ~Breit and J.A. ~Wheeler 1934 Collision of Two Light Quanta {\it Phys. Rev.} \textbf{46}  1087
\bibitem{Reiss1971} H. R. ~Reiss 1971 Production of Electron Pairs from a Zero-Mass State {\it Phys. Rev. Lett.} \textbf{26} 1072
\bibitem{Hartin2019} A. ~Hartin, A. ~Ringwald, and N. ~Tapia 2019 Measuring the boiling point of the vacuum of quantum
electrodynamics {\it Phys. Rev. D} \textbf{99}  036008
\bibitem{Blackburn2018} T.G. ~Blackburn and M. ~Marklund 2018 Nonlinear Breit-Wheeler pair creation with bremsstrahlung
$\gamma$ rays {\it Plasma Phys. Control. Fusion} \textbf{60}  054009
\bibitem{E144} C. ~Bula {\it et al.} 1996 Observation of nonlinear effects in Compton scattering {\it Phys. Rev. Lett.} \textbf{76}  3116
\bibitem{E144-2} D. L. ~Burke {\it et al.} 1997 Positron production in multi - photon light by light scattering {\it Phys. Rev. Lett.} \textbf{79}  1626
\bibitem{Cole2018} J. M. ~Cole {\it et al.} 2018 Experimental Evidence of Radiation Reaction in the Collision of a High-Intensity Laser Pulse with a Laser-Wakefield Accelerated Electron Beam {\it Phys. Rev. X} \textbf{8} 011020
\bibitem{Poder2018} K. ~Poder {\it et al.} 2018 Experimental Signatures of the Quantum Nature of Radiation Reaction in the Field of an Ultraintense Laser {\it Phys. Rev. X} \textbf{8} 31004
\bibitem{Weber2017} S. ~Weber {\it et al.} 2017 An installation for high-energy density plasma physics and ultra-high intensity laser-matter interaction at ELI-Beamlines {\it Matter and Radiation at Extremes} \textbf{2} 149
\bibitem{Gales2018} S. ~Gales {\it et al.} 2018 The extreme light infrastructure - Nuclear physics (ELI-NP) facility: New horizons in physics with 10 PW ultra-intense lasers and 20 MeV brilliant gamma beams {\it Reports on Progress in Physics} \textbf{81} 94301
\bibitem{FCC} FCC Collaboration 2021 Complementary diagnostics of high-intensity femtosecond laser pulses via vacuum acceleration of protons and electrons {\it Plasma Phys. Control. Fusion} \textbf{63} 014002
\bibitem{Jeti40} JETI40 laser system: https://www.hi-jena.de (accessed in 2021)
\end{thebibliography}
\end{document}